\documentclass[a4paper,10pt]{article}
\usepackage[utf8]{inputenc}
\usepackage{amsmath,amssymb}
\usepackage{graphicx}
\title{Effect of a non-volatile cosolvent on crack pattern induced by desiccation of a colloidal gel\footnote{Soft Matter, DOI:10.1039/C2SM25663K}}
\author{F. Boulogne, L. Pauchard, and
F. Giorgiutti-Dauphin\'e\footnote{CNRS, UMR 7608, Lab FAST, Bat 502, Campus Univ - F-91405, Orsay, France, EU. Fax: +33 1 69 15 80 60; Tel: +33 1 69 15 80 49; E-mail: pauchard@fast.u-psud.fr}}

\begin{document}

\maketitle

\begin{abstract}
 Consolidation of colloidal gels results in enormous stresses that are usually released in the formation of undesirable cracks. 
The capacity of a gel network to crack during drying depends on the existence and significance of a pressure gradient in the pore liquid;
in addition it depends on the way the gel relaxes the resulting drying stresses. 
In this paper the effect of a binary mixture of solvents saturating the gel network  on the crack patterns formation is investigated. 
Indeed, incorporation of a small quantity of non-volatile cosolvent, i.e. glycerol, inhibits drying-induced cracks; moreover addition of a concentration greater than $10\%$ to a colloidal dispersion leads to a crack free coating in room conditions. 
Mass variation with time reveals that both evaporation rate and cracking time are not affected by glycerol, in the range of added glycerol contents studied.
In addition measurements of mechanical properties show that the elastic modulus is reduced with glycerol content. 
The decrease of the number of cracks with the glycerol content is related to the flattening of the pressure gradient in the pore liquid.
The mechanism is shown to be due to the combination of two processes: flow driven by the pressure gradient and diffusion mechanisms in accordance with Scherer work (1989).
\end{abstract}

\section{Introduction}

\label{intro}

Most coatings are made by depositing a volatile liquid that contains dispersed colloidal particles on a surface. 
The liquid is then evaporated until a dry film is obtained.  
Common examples in industrial fields are paints, protective coatings (anti-corrosion), ceramic membranes \cite{Rharbi09}.
Coatings are required to be homogeneous and smooth.
Particularly they have to be free of heterogeneities possibly leading to cracking due to high stresses developing during the consolidation.
In a liquid film, evaporation of the volatile solvent causes the solid particles to be confined into a smaller volume, until they come into direct contact with each other, and form a solid porous network, so called gel \cite{Brinker90}. 
Then, further evaporation causes the liquid/air interface to curve into a set of menisci that join the particles. 
As a consequence the capillary forces exert extremely strong compressive stress on the particle network.
When the drying stress development in the gel exceeds the strength of the material, undesirable effects occur such as the formation of various modes of cracking\cite{Allain95,Lee2004,Tirumkudulu05,Xu2009}.
As a consequence the stress in the gel is relaxed.
Since numerous studies on cracks induced by drying of colloidal suspensions have been investigated for the last two decades, the stress development and the organization of colloidal particles in a drying film remain unclear.
One of the main questions that still stands: how can the stress be relaxed before cracks can form?
In addition, the development of stresses, hence of cracking notably depends on the type of the particles (size\cite{Dufresne03}, chemical composition\cite{Lee2004, Pauchard2009}) and mechanical properties of the gel formed (\cite{Atkinson91,Groisman94}), the substrate, the drying conditions (temperature\cite{Gauthier10}, relative humidity\cite{Dragnevski10}, air velocity\cite{Goehring10}) and the film thickness\cite{Atkinson91, Groisman94, Colina2000}.
In particular the capacity of a gel network to crack during drying depends on the existence and significance of a pressure gradient in the pore liquid and the way the gel relaxes the induced drying stresses. 

In the present work, we focus on the effect of cosolvent added to the colloidal dispersion on crack patterns induced by drying.
Such additive reveals crack-free drying when added to ceramic foams\cite{Fuks10}.
Glycerol has been chosen as a cosolvent for interesting properties of high miscibility in water, low volatility, and because it exhibits a highly different intrinsic diffusion coefficient from that of water.
Also glycerol is an element composing many essential fine arts products\cite{Gettens1966}: due to its transparency properties and because it is used as a softener, avoiding creases and cracks.
Crack patterns have been quantified as a function of the added glycerol content.
Drying kinetics have been investigated by mass variation with time measurements.
Also mechanical properties of colloidal gels are measured by indentation testing as a function of the added cosolvent content.
Experimental results can be explained by the combination of two processes occuring during the drying process: flow driven by the pressure gradient accordingly with Darcy law and diffusion mechanism accordingly with Fick law.

\section{Experimental}
\label{experimental}

\subsection{Materials and samples preparation}

\begin{figure}
\centering
  \includegraphics{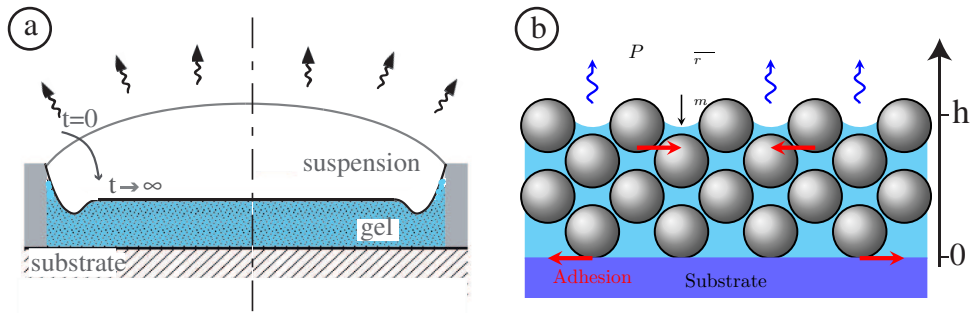}

\caption{(a) Experimental set-up (side view): during solvent loss the layer consolidates. (b) Sketch of the solid network saturated by solvent: curvature of the solvent/air menisci occurs at the evaporation surface during solvent loss and exerts an enormous stress on this network.}
\label{fig:setup}       
\end{figure}

We used colloidal silica dispersion, Ludox HS-40 purchased from Sigma-Aldrich.
The radius of the particles is $9 \pm 2$ nm, the density $2.36\times 10^3$ kg/m$^3$.
The mass fraction of the initial dispersion was estimated to $\phi_m = 40.5$\% by a dry extract. 
Thus, we could deduce the particle volume fraction for this solution to be $\phi_v \simeq 22$\%.
Particles are polydisperse enough not to crystallize.
In the absence of evaporation, the suspension is stable due to the competition between the attractive Van der Waals forces and the electrostatic repulsion\cite{Laven2001} between particles (pH = $9.8$).

Glycerol (purity: $99.0$\%, from Merck) was added to the aqueous dispersion as a cosolvent.
This solvent exhibits high miscibility properties in water and low vapor pressure ($2.2\times10^{-4}$ mbar at $25^\circ$C)\cite{Daubert1989} compared to water ($31.7$ mbar). 
Thus, we assume glycerol as a non-volatile compound.
The surface tension of pure glycerol is $62.5$mN/m at $20^{\circ}$C \cite{Lide08}.
Glycerol/water weight ratio was prepared using the following process:
The resulting solution was filtered to $0.5$~$\mu$m to eliminate eventual aggregates; the mass fraction is estimated by a dry extract at $140^\circ$C. 
In order to keep the particle mass fraction constant for each sample, a given quantity of water is evaporated, then replaced by an equal mass of glycerol;
the evaporation of water is achieved by stirring the aqueous dispersion at $35^\circ$C in an Erlenmeyer.  
Finally, we checked the final pH of the solution. 

We note $\kappa$ the ratio between the mass of glycerol added and the mass of the resulting dispersion. 
Six solutions have been investigated: $\kappa=1.1$, $2.1$, $3.1$, $5.1$, $7.2$ and $8.9$ $\%$ wt. 
This leads to glycerol fraction in solvent denoted by $\chi_i$ (the subscript stands for initial).
In the range of glycerol content added, the solvent viscosity increases until $1.3$ mPa.s.

\subsection{Drying geometry}

Experiments are performed on films of fixed thickness.
Also a small amount of the solution (initial weight $m_i = 0.52\pm0.02$ g) is deposited inside a circular container (diameter $= 15$ mm) whose bottom is a glass microscope slide carefully cleaned with ethanol and the lateral wall is a rigid rubber ring (figure \ref{fig:setup} (a)).
The contact line of the solution is quenched at the upper edge of the circular wall and remains there during the whole experiment. 
In this geometry the layer dries at room temperature ($T=23^\circ$C) under controlled relative humidity ($R_H \sim 50 \pm 2 \%$).
In this way, the desiccation takes place in the absence of convection in the vapor so that the evaporation is limited by diffusion of the solvent into the air. 
The drying kinetics is obtained using a scale (Sartorius) with an accuracy of $0.01$ mg.
When the gelled phase is formed, the layer exhibits an approximately constant thickness in a central region (figure \ref{fig:setup} (a)). 
In this region, covering about $70\%$ of the total surface area, the evaporation is assumed to be uniform.
A typical value of the thickness far from the border is $h \sim 0.7$ mm.
The colloidal gel is transparent allowing us to observe easily the dynamics of crack patterns in the layer.
Also the crack patterns formation is recorded using an interval timer with an AVT's Marlin camera positioned on the top of the sample.

\subsection{Measurements of the mechanical properties of the gel phase}

To characterize the mechanical properties of the gelled layers relevant elastic modulus, $E_{p}$, has been performed by indentation testing.
This quantity has been determined using a CSM Instruments Micro Indentation Tester (MHT) with a four-sided pyramidal Vickers indenter.
Indentation testing is a local method of mechanical investigation.
The indenter, initially in contact with the surface of the gel, is driven in the material until a maximal $70$ mN load with a loading speed $70$ mN/min. The maximal force is held during $10$ s.
Then, the load is decreased until zero with the same speed.
Figure \ref{fig:colloid_E}a shows a typical load/unload curve.
The standard way of estimating the elastic modulus from the indentation load-displacement curve uses the initial slope of the unloading curve.
Based on Sneddon work \cite{Sneddon65}, the following formula has been derived for the elastic modulus, $E_{p}$, to the unloading slope \cite{Oliver1992,Malzbender2002}:

$$E_{p} = \frac{\sqrt{\pi}}{2} \frac{S}{\sqrt{A}}$$

In this equation, $S$ is the slope of the unloading curve at the start of unloading, and $A$ is the projected area of contact between the indenter and the material at that point (inset in figure \ref{fig:colloid_E}a). 
Note that the contact between the indenter and the gel is always chosen in a location where the layer adheres on the substrate.
In the case of successive measurements of the elastic modulus with time, indentation testing was performed in the same adhering region to obtain comparable values as effectively as possible.
Particularly, the difficulties of these measurements lie in gel films without glycerol addition exhibiting the smallest size of adhering regions.
Indeed it has been observed that glycerol reduces the capacity of the gel to delaminate.
In addition indentation testing applied to drying gelled layers has to be carefully analyzed since the system is not homogeneous in the thickness and evolves with time.
Also measurements of elastic moduli in gels by this way do not give absolute values but provide on the one hand a process of comparing mechanical properties of different materials, on the other hand a characterization of the time evolution of the mechanical properties of the system.
Therefore measurements for the elastic modulus have been reproduced several times and performed on 2 or 3 samples;
bars in figure \ref{fig:colloid_E}b take into account the minimum and maximum measurements obtained.

\section{Results}
\label{Results}

\subsection{Drying}

\label{drying}
\begin{figure}
\centering
\resizebox{1\columnwidth}{!}{
  \includegraphics{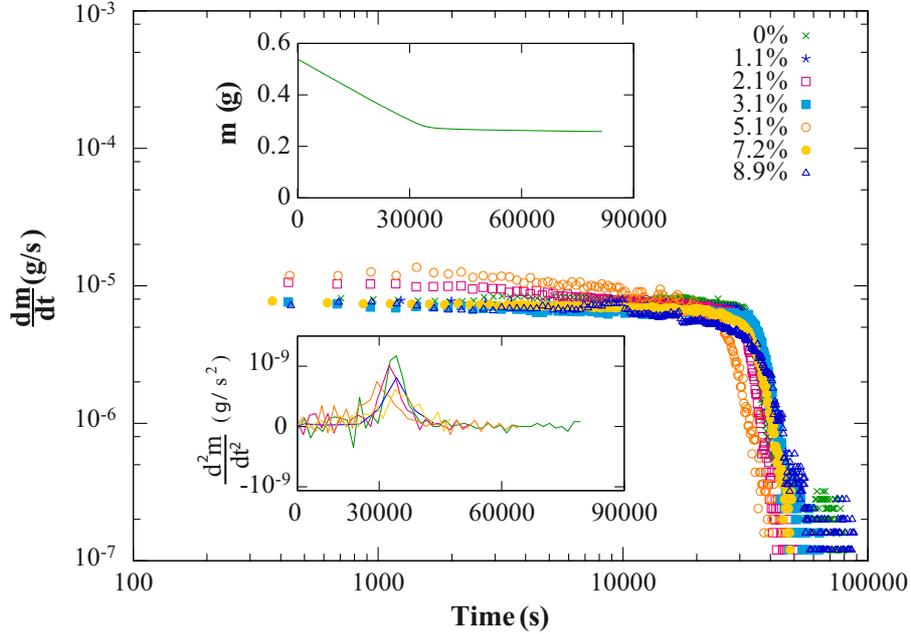}
}
\caption{Drying curves. Drying rate, $\frac{dm}{dt}$, for silica films for different additional glycerol concentrations, $\kappa$. The drying rate is calculated from the mass variation with time (upper inset). The variation $\frac{\partial^2 m}{\partial t^2}$ with time evidences the transient time close to $32000$s.}
\label{fig:masse}       
\end{figure}

Once the layer is deposited in the circular trough, evaporation of the water takes place.
The drying process of a colloidal dispersion is usually separated into two
stages, that can be roughly distinguished by measuring mass, $m$, and drying
rate, $\frac{dm}{dt}$, variations with time (figure \ref{fig:masse})\cite{Scherer1987}. 
In the first regime (constant rate period) the evaporation rate is constant and evaporation is mainly controlled by the external conditions of relative humidity and temperature in the surroundings.
During this stage, water removal concentrates the colloidal particles into a closed packed array: a porous network saturated with solvent.
Due to geometry of the planar film, evaporation at the gel-air interface and adhesion at the gel-substrate interface result in the development of high stresses in the gel (figure \ref{fig:setup}(b)).
Capillary pressure that is caused by the liquid menisci formed between the particles at the top of the packed region is responsible for the shrinkage of the porous network.
The shrinkage of the gel at the gel-air interface is frustrated by the adhesion to the substrate. 
As a consequence tensile stresses progressively build up in the layer. 
Then a non-linear drying rate follows: the falling rate period, when liquid/air interfaces recede into the porous medium and when the drying process is limited by flow of liquid to the top layer through the porosity. 
The transient between the constant rate period and the falling rate period is highlighted in the lower inset in Figure \ref{fig:masse}: it shows the drying rate derivative $\frac{d^2m}{dt^2}$. 
In our experimental conditions we observe that the transient always occurs at $\sim32000$s in the studied $\kappa$ range. 
Consequently the drying rate decreases as clearly shown in figure \ref{fig:masse}. 
Similar behaviors occur when glycerol content is added to the aqueous dispersion (figure \ref{fig:masse}).

\subsection{Cracking}
\label{cracking}

\begin{figure}
\centering
\resizebox{.8\columnwidth}{!}{
  \includegraphics{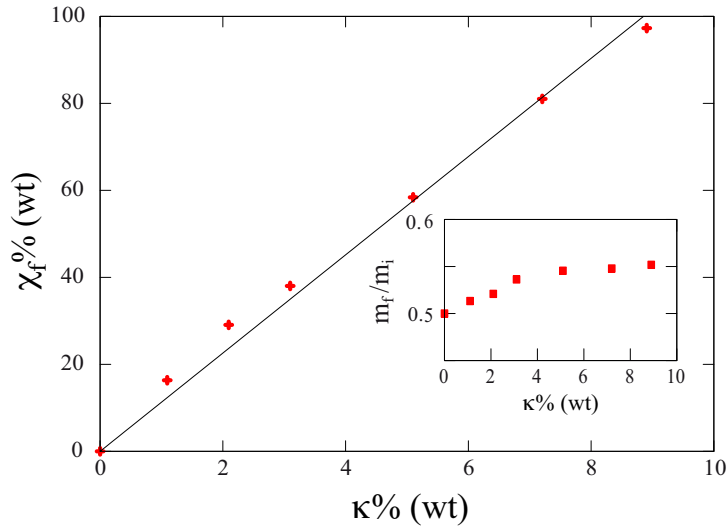}
}
\caption{Mass ratio of samples for different glycerol concentration $\kappa$. The inset shows the final concentration of glycerol in the solvent wetting the porous medium as a function of $\kappa$. The solid line is a linear adjustment (slope$ = 11.3\pm0.3$).}
\label{fig:masse_fin}       
\end{figure}

As the tensile stress reaches a threshold value, cracks appear in the film and invade the plane of the layer.
First cracks appear at $t\in[24000,27000]$s after deposition of the layer on the substrate, without any correlation with the glycerol concentration nor the initial mass disparity. 
Moreover the cracks invade the layer during only a few minutes.
Thus the crack network develops during the initial stage of the drying process, in constant rate period \ref{fig:masse}, and does not evolve afterwards.
Plot in inset in figure \ref{fig:masse_fin} shows the ratio, $m_{f}/m_{i}$, of the final (at $t = 60000$s) and the initial mass of material on the glass plate (raw data).
Futhermore, we know that silica particles contributes to $40$\% of the mass.
Thus, the excess of mass is due to the solvent.
As previously mentioned only water could evaporate and glycerol remains in the porous medium. 
Also we can calculate the mass fraction of glycerol $\chi_f$ composing the solvent when the mass variations with time do not evolve anymore ($t \sim 60000$s):

\begin{equation}
\label{e.1}
 \chi_f = 1- \frac{m_f/m_i - 0.4 - \kappa}{m_f/m_i - 0.4 }
\end{equation}

We found that the higher the initial fraction $\kappa$ is, the higher $\chi_f$ reaches (see inset of figure \ref{fig:masse_fin}), leading to a saturation of glycerol in the gel for $\kappa > 8.9 \pm 0.3 \%$ in accordance with the fit. 

\begin{figure}
\centering
\resizebox{1\columnwidth}{!}{
  \includegraphics{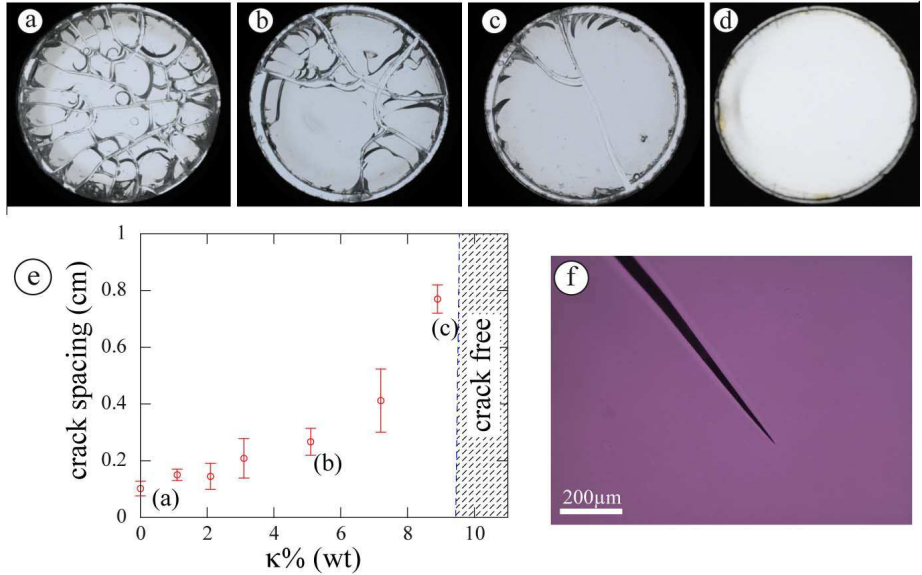}

}\caption{Crack patterns after 16 hours in gelled layers when adding a second solvent to the system. (a) Without any additional glycerol. (b) $\kappa = 5.1\%$. (c) $\kappa = 8.9\%$.
(d) $\kappa = 10.2\%$: the layer is crack-free at the mesoscopic scale. (e) Crack spacing plotted as a function of the concentration in glycerol (drying conditions $R_H = 50 \%$ and $T = 23^\circ$C; the initial weight deposited in each trough is the same). (f) Propagating crack tip in a gel with additional glycerol $\kappa = 8.9\%$.}
\label{fig:patterns}       
\end{figure}

Crack patterns are strongly modified by the presence of a glycerol within the solvent.
Indeed, in the case of a silica gel without additional glycerol, the successive formation of cracks results in the typical pattern shown in figure \ref{fig:patterns}a.
This pattern is usually continued by further crack generations: particularly delamination process takes place when the gel detaches from the substrate \cite{Pauchard06}.

Moreover final crack patterns are shown in figure \ref{fig:patterns}a,b,c,d for increasing glycerol concentrations $\kappa$ up to $10.2\%$ in our experimental conditions.
For each pattern, cracks still interconnect leading to complete network: no broken network is observable suggesting that if crack nucleation is less favourable in presence of additional glycerol, crack propagation is still not inhibited.
The number of cracks generated diminishes when the quantity of additional glycerol increases; this results in increasing the crack spacing as shown in figure \ref{fig:patterns}e.
Figure \ref{fig:patterns}f shows a magnification of a typical crack tip propagating in a gel with glycerol content.
The sharpness of this crack tip, also observed for every $\kappa \lesssim 9.5\%$, suggests that the gel still exhibits brittleness properties. 
In addition buckle-driven delamination process, that usually takes place after the formation of a crack network, is inhibited by increasing additional glycerol concentration in the gel. 

Above a threshold concentration of additional glycerol, the layer remains crack free (figure \ref{fig:patterns}d): no crack forms because the gel could not reach the compaction (or the threshold stress) for which cracks propagate. 
It results in a flat gel film saturated in glycerol.

\subsection{Gel consolidation}
\label{Gel consolidation}

\begin{figure}
\centering
  \includegraphics[scale=0.7]{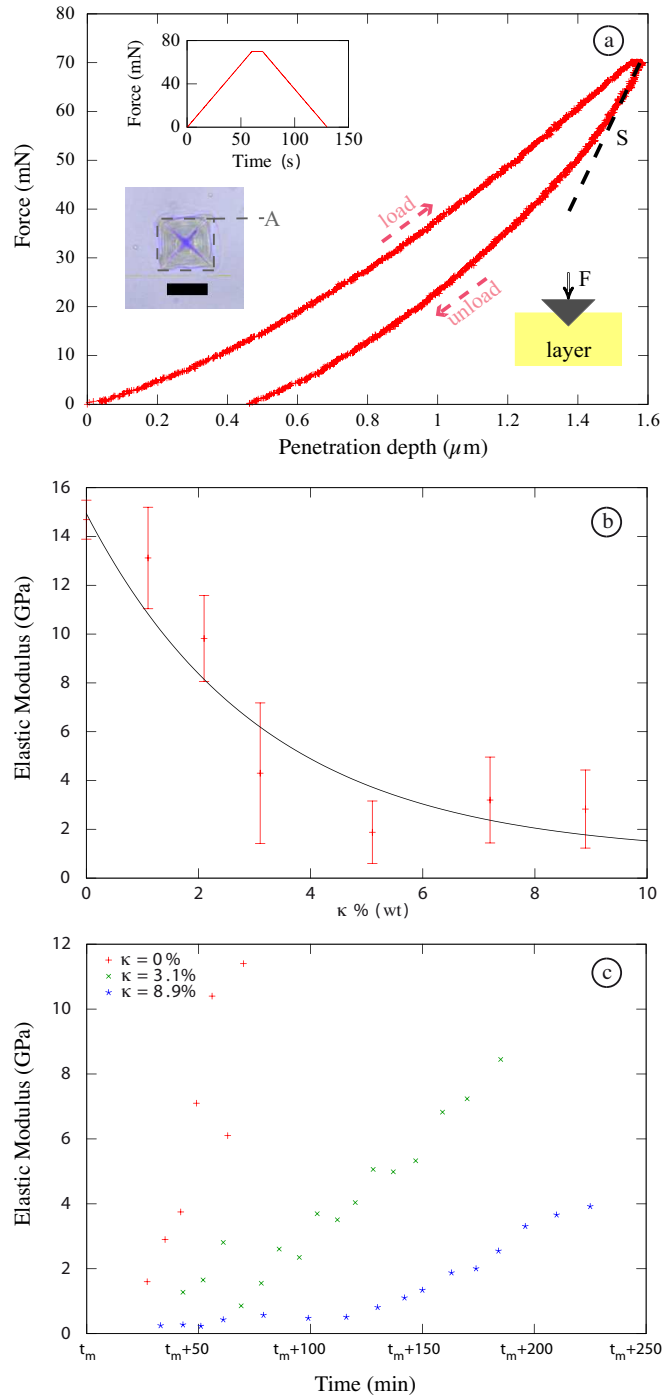}
\caption{Macroscopic elastic response of the gel obtained by indentation testing. (a) Loading/unloading curve: force applied on the tip versus indenter displacement ($\mu m$). Insets: variation of the applied force versus time; image of indentation print at the surface of the gel after a loading/unloading cycle observed using optical microscopy (maximum load: $70 mN$, bar $= 25\mu m$): the dashed square limit the projected area of contact, $A$. (b) Variation of the elastic modulus, $E_{p}$, for seven glycerol-Ludox solutions. 
(c) Variation of the elastic modulus, $E_{p}$, as a function of time for Ludox HS-40 without glycerol addition ($\kappa=0\%$) and for two different concentrations of glycerol ($\kappa=3.1\%,8.9\%$). Measurements start after crack propagation, $t>t_{m}$.
}
\label{fig:colloid_E}       
\end{figure}

The macroscopic elastic response of the gel is characterized by indentation testing for various binary mixtures of solvents (figure \ref{fig:colloid_E}).
Based on the procedure depicted in figure \ref{fig:colloid_E}a, elastic moduli have been measured for gels of various glycerol contents.
Each measurement has been obtained $\sim 8$ hours after the cracks formation since the gel evolves slowly after this duration (figure \ref{fig:masse}).
Measurements in figure \ref{fig:colloid_E}b reveal that an increase of the quantity of glycerol in the gel results in a decrease of its elastic modulus.
Since glycerol remains inside the porous medium, it could act as a wetting agent lowering the material stiffness.

The elastic modulus of silica gels has been measured as a function of time during the drying process; measurements started after crack propagation, $t>t_{m}$.
These measurements are shown for the silica gel without any glycerol addition and for two glycerol contents in figure \ref{fig:colloid_E}c.
In any cases, gel films clearly stiffens as they consolidate with time; for the silica gel without any glycerol addition, this process is more rapid than in presence of glycerol content.
Two factors could contribute to increase the modulus as a function of time: stiffening of the solid phase of the gel and reduction in the porosity \cite{Scherer89bis}.
Also the elastic modulus increases faster for gels having lower concentrations of glycerol.


\section{Discussion} 
The capacity of a gel to crack during drying depends on the existence and significance of a pressure gradient in the pore liquid.
Indeed the spatial variation in pressure causes a variation in contraction of the gel network \cite{Scherer89}.
This modifies the way the drying stresses are relaxed and consequently the crack are formed.
In the following the spatial variation in pressure in the liquid pore is characterized in gels saturated by binary mixtures of solvents during the first drying regime since cracks form during this stage.
Particularly we have chosen to compare the pressure distribution for different glycerol contents at time $t_{m} = 25500$s. 
At this moment, close to the typical cracking time of the gelled layers, gels can be considered as non-shrinking, that is the gel network does not change during time\cite{Scherer89}; pores contain the binary mixture of solvents and the liquid-air interfaces are already curved.
 
 In drying gels, liquid usually moves through the network in response to a pressure gradient of liquid, $\frac{\partial P}{\partial z}$, accordingly to the Darcy law\cite{Brinker90}.
Also, in the case of a gel saturated with a single volatile solvent, the flux of evaporating liquid at the gel-air interface comes entirely from flow of liquid. 
It expresses as:

\begin{equation}
\label{e.2}
J_{Darcy} = \frac{k}{\eta}\frac{\partial P}{\partial z}
\end{equation}

where $k$ is the network permeability of the gel, and $\eta$ is the dynamic viscosity of the solvent in the pore; note that the pressure of the liquid pore is regarded as positive.

Let us now consider the incorporation of a soluble and non-volatile cosolvent, i.e. glycerol, to the aqueous dispersion.
The drying process leads to a gel phase saturated by a binary mixture of water and glycerol.
This results in a combination of the pressure flow component (\ref{e.2}) and a diffusion mechanism\cite{Crank75,Scherer89,Brinker90}.
In that way, the diffusive flux for each component, $i$, expresses in accordance with the Fick law:

\begin{equation}
\label{e.3}
J_{Fick}^{i} = -D_{i}\frac{\partial C_{i}}{\partial z}
\end{equation}

where $D_{i}$ is the intrinsic diffusion coefficient refers in terms of the rate of transfer of component $i$ across a fixed section.
Note that $P$ and $C_{i}$ are time dependent.

The total fluid flux in the porous network combines the flow driven by the pressure gradient (\ref{e.2}) and the diffusive flux (\ref{e.3}).
Also, the relative change in volume of liquid over time, $\dot{V_{L}}/V_{L}$, is due to the change in both local flow, $\partial_{z} J_{Darcy}$, and local diffusion flux, $\partial_{z} J_{Fick}^i$.
As a result the following conservation equation can be written\cite{Scherer89}:

\begin{equation}
\label{e.4}
(1-\rho)\frac{\dot{V_{L}}}{V_{L}} = - \partial_{z} J_{Darcy} -  K \partial_{z} (J_{Fick}^1+J_{Fick}^2)
\end{equation}

where $\rho$ is the relative density of the gel, $K$ denotes the porosity of the gel, $J_{Fick}^i$ is the diffusive flux of component $i$, stating $i=1$ for water and $i=2$ for glycerol.

Since the change in the volume of liquid has to equal the change in pore volume, the following relation can be written in accordance with reference\cite{Scherer89}:

\begin{equation}
\label{e.5}
(1-\rho)\frac{\dot{V_{L}}}{V_{L}} =  - \frac{1}{E_{p}}  \left(\beta  \dot{P} + (1-\beta)<\dot{P}>\right)
\end{equation}

where $\beta$ is a function of the Poisson ratio of the gel and $\dot{P}$ denotes the time derivative of the pressure.
Also the flux from the gel surface equals the rate of change in volume of the gel; this expresses as $<\dot{P}> = E_{p}\frac{V_{E}}{h}$ for a layer of thickness $h$ accordingly to reference\cite{Scherer89}.
Here the evaporation rate, $V_{E}$, determines the rate of volume loss by water evaporation.

By equating the right sides of expressions \ref{e.4} and \ref{e.5} and taking into account relations \ref{e.2} and \ref{e.3}, it comes: 

\begin{equation}
\label{e.6}
\frac{\beta }{E_{p}}  \frac{\partial P}{\partial t} + (1-\beta)\frac{V_{E}}{h}=  \frac{k}{\eta} \frac{\partial^2 P}{\partial z^2} + K \left(D_2-D_1\right) \frac{\partial^2 C_{1}}{\partial z^2}
\end{equation}

Equation \ref{e.6} reveals that the diffusion mechanism can become significant only when the intrinsic diffusion coefficients of cosolvents are sufficiently different from each other.
Note that equation \ref{e.6} is only valid in the case of a solvent made of two components: in the case of a pure solvent, the diffusive term (second term of the right side) disappears and the flow is only governed by the Darcy law.
Applying equation \ref{e.6} to our system allows us to plot the pressure in a pore liquid of the gel for different glycerol contents.
In that purpose the concentration of component 1 (water), $C_{1}$, has to be incorporated in equation \ref{e.6}; $C_{1}$ satisfies the diffusion equation using the boundary condition of constant flux at the evaporation surface, that expresses as (reference\cite{Crank75} p. 61):
$-(\bar{C_{1}}D_{1}+(1-\bar{C_{1}})D_{2})  \frac{\partial C_{1}}{\partial z}\mid_{z=h} = V_{E}$, 
where $\bar{C_1}$ is the mean concentration of water in the layer defined as: $\bar{C_1}(t_{m}) = \frac{1}{h}\int_0^h C_{1}(z,t_{m}) dz$ at time $t_{m}$.

\begin{figure}
\centering
\resizebox{1\columnwidth}{!}{
  \includegraphics{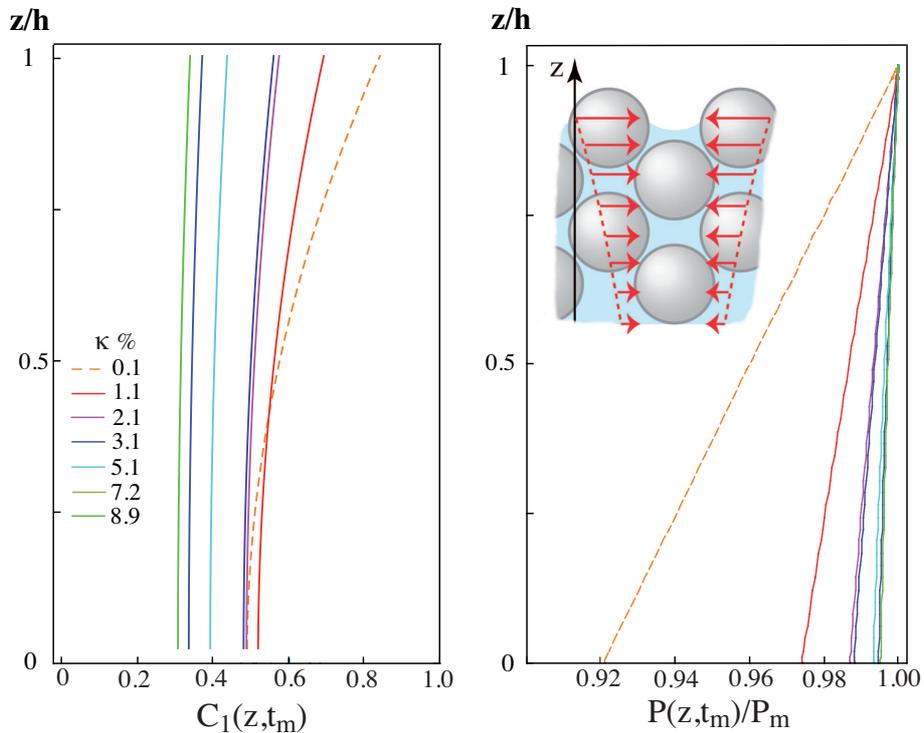}
}
\caption{Concentration distribution in water, $C_{1}$, (left) and normalized pressure distribution, $P/P_{m}$, (right) along the pore liquid, for several initial glycerol contents, $\kappa$. These quantities are plotted at time $t_{m}$ when the highest pressure value reaches $P_{m}$ at the evaporation surface. Inset: sketch of the pressure gradient in the liquid pore.}
\label{fig:scherer}       
\end{figure}

As a result an estimation of each quantity of equation \ref{e.6} from the experimental results is required.
$\bar{C_1}(t_m)$ is precisely estimated from the inset in figure \ref{fig:masse_fin}.
In the first drying regime, $V_{E}$ can be measured from the mass variation with time in figure \ref{fig:masse} ( $V_E \simeq 4 \times 10^{-8}$ m/s).
The values of the intrinsic coefficients are obtained from reference \cite{Vitagliano04}: $\frac{D_{2}}{D_{1}} \sim 50$.
The porosity $K$ is roughly estimated to $0.40$ for the random close-packed phase while $\beta = 1/2$ (reference \cite{Scherer89}).
We assume that the permeability of the gel keeps a constant value during the crack formation; this quantity is estimated using the Carman-Kozeny relation ($k = 4.9 \times 10^{-19}m^2$)\cite{Dullien91}.
Finally the elastic moduli of gels composed with various glycerol contents have been estimated from experimental results shown in figure \ref{fig:colloid_E}(b) and (c) extrapolating values a short time after cracking.
Consequently, the concentration distribution $C_{1}$ and the pressure, $P$, are plotted along liquid pores at time $t_{m}$ in figure \ref{fig:scherer} for different glycerol contents.
The flattening of the pressure gradient when the concentration of glycerol increases is clearly observable and is mainly attributed to the diffusion mechanism. One should note that the reduction in the capillary forces near the surface should be of minor importance as the surface tension from one mixture to another varies only from  $\sim 68$mN/m to $\sim72$mN/m \cite{Lide08}. 
These results highlight the effect of the addition of a non-volatile solvent to the pressure in the liquid pore since diffusion can extract liquid from the interior and flatten the pressure gradient in the pore liquid.
The pressure gradient causes a shrinkage that is transmitted in the hard particles network.
In the present paper, the adhesion with the substrate is assumed to be a constant and to play a determinant role as responsible for the development of high stresses in the gel\cite{CimaChiu93,Cima93}.
Nevertheless the adhesion is seen as a necessary but not sufficient condition for cracking.
Thus we eliminate cracking by modifying the pressure gradient and keeping constant the adhesion.
The increase of glycerol content, through diffusion process, will imply that the resulting drying stresses will be distributed more uniformly in the gel.
Consequently, the stress concentration in the gel network is limited and the crack formation can be inhibited.

The effect of addition of glycerol content results in the flattening of the pressure gradient in the pore liquid due to a diffusion process as depicted by equation \ref{e.6}, for a constant value of the elastic modulus.
Moreover, the effect of addition of glycerol content results in the tendency of a decrease of the elastic modulus of the gel.
As a consequence a decrease of the elastic modulus in equation \ref{e.6} leads to a flattening of the pressure gradient, and, in that sense, emphasizes the effect of diffusion.

\section{Conclusion}
The incorporation of a small percentage of non-volatile cosolvent modifies the way the drying stresses relaxe in a colloidal gel.
Experimental data show that the drying rate and cracking time is not significantly modified by incorporation of the cosolvent. 
However the presence of non-volatile decreases the stiffness of the gel at a given consolidation time and inhibits crack formation:
incorporation of a concentration of glycerol greater than $10\%$ in the colloidal dispersion results in a crack free drying gel.
The mechanism is shown to be due to the combination of both flow driven by the pressure gradient and diffusion of solvent into a another; in this scenario the diffusion induced by evaporation causes a demixing effect of solvents.
This results in a flattening of the pressure gradient in the pore liquid and consequently the distribution of the drying stresses is more uniform, limiting stress concentration in the gel network possibly cause of cracking.
Moreover, the effect of a cosolvent added to a silica dispersion on the crack pattern has been highlighted in the case of glucose.
As in the case of glycerol, the presence of glucose in the gel film reveals an increase of the crack spacing. 
However, contrary to the case of glycerol, cracks are not completely inhibited above a threshold glucose content: the crack spacing tends to a maximum value (typically $0.45$cm in the same experimental conditions as for glycerol cosolvent). 

Finally, the effect of the flattening of the pressure gradient is surely not the only process leading to the diminution of the number of cracks in the gel film.
Particularly chemical effects between glycerol and silanol groups covering the surfaces of silica particles could change the strength of the gel film.
This process is suggested by aggregation rate measurements of silica particles with polyols\cite{Gulley2001} and would be of interest for further works applied to gel film cracks induced by desiccation with chemical additives.

\footnotesize{
\bibliography{cracks_glycerol} 

\begin{thebibliography}{10}

\bibitem{Allain95}
C.~Allain and L.~Limat.
\newblock {\em Phys. Rev. Lett.}, 74:2981, 1995.

\bibitem{Atkinson91}
A.~Atkinson and R.~M. Guppy.
\newblock {\em J. Mater. Sci.}, 26:3869--3873, 1991.

\bibitem{Rharbi09}
N.~Bassou and Y.~Rharbi.
\newblock Role of b\'enard-marangoni instabilities during solvent evaporation
  in polymer surface corrugations.
\newblock {\em Langmuir}, 25(1):624--632, 2009.

\bibitem{Brinker90}
C.~J. Brinker and G.~W. Scherer.
\newblock {\em Sol-Gel Science: The Physics and Chemistry of Sol-Gel
  Processing}.
\newblock Elsevier Science, 1990.

\bibitem{CimaChiu93}
R.~C. Chiu, T.~J. Garino, and M.~J. Cima.
\newblock Drying of granular ceramic films: I, effect of processing variables
  on cracking behavior.
\newblock {\em Journal of the American Ceramic Society}, 76(9):2257--2264,
  1993.

\bibitem{Cima93}
Raymond~C. Chiu and Michael~J. Cima.
\newblock Drying of granular ceramic films: Ii, drying stress and saturation
  uniformity.
\newblock {\em Journal of the American Ceramic Society}, 76(11):2769--2777,
  1993.

\bibitem{Colina2000}
H.~Colina and S.~Roux.
\newblock Experimental model of cracking induced by drying shrinkage.
\newblock {\em The European Physical Journal E: Soft Matter and Biological
  Physics}, 2000.

\bibitem{Crank75}
J.~Crank.
\newblock {\em The Mathematics of Diffusion}.
\newblock Clarendon Press, 2nd ed edition, 1975.

\bibitem{Daubert1989}
T.~E. Daubert and R.~P. Danner.
\newblock {\em Physical and thermodynamic properties of pure chemicals: data
  compilation}.
\newblock Hemisphere Publishing Corp New York, 4 edition, 1989.

\bibitem{Vitagliano04}
G.~D'Errico, O.~Ortona, F.~Capuano, and V.~Vitagliano.
\newblock {\em J. Chem. Eng. Data}, 49:1665--1670, 2004.

\bibitem{Dragnevski10}
K.~I. Dragnevski, A.~F. Routh, M.~W. Murray, and A.~M. Donald.
\newblock {\em Langmuir}, 26:7747--7751, 2010.

\bibitem{Dufresne03}
E.~R. Dufresne, E.~I. Corwin, N.~A. Greenblatt, J.~Ashmore, D.~Y. Wang, A.~D.
  Dinsmore, J.~X. Cheng, X.~S. Xie, J.~W. Hutchinson, and D.~A. Weitz.
\newblock {\em Phys. Rev. Lett.}, 91:224501, 2003.

\bibitem{Dullien91}
F.A.L. Dullien.
\newblock {\em Porous Media, Second Edition: Fluid Transport and Pore
  Structure}.
\newblock Academic Press, 1991.

\bibitem{Fuks10}
D.~Fuks, G.~E. Shter, M.~Mann-Lahav, and G.~S. Grader.
\newblock Crack-free drying of ceramic foams by the use of viscous cosolvents.
\newblock {\em Journal of the American Ceramic Society}, (93):3632--3636, 2010.

\bibitem{Gauthier10}
G.~Gauthier, V.~Lazarus, and L.~Pauchard.
\newblock {\em Europhys. Lett.}, 89:26002, 2010.

\bibitem{Gettens1966}
Rutherford~J. Gettens and George~L. Stout.
\newblock {\em Painting materials : a short encyclopaedia}.
\newblock Peter Smith Pub, 1966.

\bibitem{Goehring10}
L.~Goehring, W.~J. Clegg, and A.~F. Routh.
\newblock {\em Langmuir}, 26:9269 -- 9275, 2010.

\bibitem{Groisman94}
A.~Groisman and E.~Kaplan.
\newblock {\em Europhys. Lett.}, 25:415--420, 1994.

\bibitem{Gulley2001}
Gerald~L. Gulley and James~E. Martin.
\newblock Stabilization of colloidal silica using polyols.
\newblock {\em Journal of Colloid and Interface Science}, 241(2):340--345,
  2001.

\bibitem{Laven2001}
Jozua Laven and Hans~N. Stein.
\newblock The electroviscous behavior of aqueous dispersions of amorphous
  silica (ludox).
\newblock {\em Journal of Colloid and Interface Science}, 2001.

\bibitem{Lee2004}
W.~P. Lee and A.~F. Routh.
\newblock Why do drying films crack?
\newblock {\em Langmuir}, 20(23):9885--9888, 2004.
\newblock PMID: 15518466.

\bibitem{Lide08}
D.R. Lide.
\newblock {\em CRC Handbook of Chemistry and Physics}.
\newblock CRC Press/Taylor and Francis, 89th edition edition, 2008.

\bibitem{Malzbender2002}
J.~Malzbender, J.~M.~J. den Toonder, A.~R. Balkenende, and G.~de~With.
\newblock Measuring mechanical properties of coatings: a methodology applied to
  nano-particle-filled sol-gel coatings on glass.
\newblock {\em Materials Science and Engineering: R: Reports}, 2002.

\bibitem{Oliver1992}
W.C. Oliver and G.M. Pharr.
\newblock {\em J. Mater Res.}, 1992.

\bibitem{Pauchard06}
L.~Pauchard.
\newblock Patterns caused by buckle-driven delamination in desiccated colloidal
  gels.
\newblock {\em Europhys. Lett.}, 74(1):188--194, 2006.

\bibitem{Pauchard2009}
L.~Pauchard, B.~Abou, and K.~Sekimoto.
\newblock Influence of mechanical properties of nanoparticles on macrocrack
  formation.
\newblock {\em Langmuir}, 25(12):6672--6677, 2009.

\bibitem{Scherer89bis}
G.~W. Scherer.
\newblock {\em Journal of Non-Crystalline Solids}, 109:183--190, 1989.

\bibitem{Scherer89}
G.~W. Scherer.
\newblock {\em Journal of Non-Crystalline Solids}, 107:135--148, 1989.

\bibitem{Scherer1987}
George~W. Scherer.
\newblock Drying gels: V. rigid gels.
\newblock {\em Journal of Non-Crystalline Solids}, 1987.

\bibitem{Sneddon65}
I.~N. Sneddon.
\newblock The relation between load and penetration in the axisymmetric
  boussinesq problem for a punch of arbitrary profile.
\newblock {\em Int. J. Eng. Sci.}, 3:47--57, 1965.

\bibitem{Tirumkudulu05}
M.~S. Tirumkudulu and W.~B. Russel.
\newblock {\em Langmuir}, 21:4938, 2005.

\bibitem{Xu2009}
Peng Xu, A.~S. Mujumdar, and B.~Yu.
\newblock Drying-induced cracks in thin film fabricated from colloidal
  dispersions.
\newblock {\em Drying Technology: An International Journal}, 27(5):636--652,
  2009.

\end{thebibliography}
\bibliographystyle{plain} 
}

\section{Acknowledgment}
The authors thank A. Aubertin, L. Auffray, C. Borget and R. Pidoux for technical supports.
In addition we thank the referees for their valuable comments which improved the consistancy of the manuscript.

\end{document}